\documentclass[aps,prc,twocolumn,superscriptaddress]{revtex4-1}
\usepackage{graphicx}
\usepackage{amsmath}
\usepackage{subfigure}
\usepackage{subfigure,dcolumn}
\usepackage{appendix}
\usepackage{xcolor}
\usepackage{ulem}
\usepackage[colorlinks,
linkcolor=blue,
anchorcolor=blue,
urlcolor=red,
citecolor=blue]{hyperref}
\begin{document}
	
	
	
\title{A novel filtering method for generating desired density profiles of colliding nuclei}
	

\author{Xilong Xiang}
\affiliation{College of Physics Science and Technology, Shenyang Normal University, Shenyang 110034, China}
\affiliation{School of Science, Huzhou University, Huzhou 313000, China}
\author{Manzi Nan}
\affiliation{School of Science, Huzhou University, Huzhou 313000, China}
\author{Pengcheng Li}
\email[Corresponding author, ]{lipch@zjhu.edu.cn}
\affiliation{School of Science, Huzhou University, Huzhou 313000, China}
\author{Yongjia Wang}
\affiliation{School of Science, Huzhou University, Huzhou 313000, China}
\author{Ling Liu}
\affiliation{College of Physics Science and Technology, Shenyang Normal University, Shenyang 110034, China}
\author{Qingfeng Li}
\affiliation{School of Science, Huzhou University, Huzhou 313000, China}

\date{\today}

\begin{abstract}		

Accurate modeling of the density profile is essential for studying heavy-ion collisions (HICs) with a transport model. Within the framework of the quantum molecular dynamics (QMD)-type model, a novel method for generating desired nuclear density distributions based on Fourier series expansion is proposed. This new initialization method is further incorporated into the ultrarelativistic quantum molecular dynamics model, and the bubble-like density distribution of $^{96}$Ru is constructed. Then, by simulating $^{96}$Ru+$^{96}$Ru collisions at
$E_\mathrm{lab}=1500~\mathrm{MeV/nucleon}$ with different equations of state (EoS) and initialization methods, the effects of the initial density distribution on the final state observables and the constrained information of EoS are analyzed. It is found that $^{96}$Ru nuclei with a bubble density profile lead to an increased maximum compression during the collision, which in turn enhances the collective flow. Moreover, a relatively stiff EoS with $K_0>280$ MeV is favored for the conventional Woods–Saxon type density profile, whereas an EoS with $K_0$=200–280 MeV is supported when a bubble-like density profile is employed. These results demonstrate that the initial nuclear density distribution plays a non-negligible role in dynamical observables and EoS constraints. The proposed method thus provides a powerful tool for constructing exotic profiles and investigating nuclear structure effects in HICs.

\end{abstract}

\pacs{}

\maketitle
	
\section{Introduction}\label{sec:1}

The precise constraint of the nuclear equation of state (EoS) at high densities is one of the current frontier issues in nuclear physics and nuclear astrophysics. It plays an important role in understanding nuclear structure, heavy-ion reactions, as well as the formation and evolution of compact stars \cite{Danielewicz:2002pu,Colonna:2020euy,Huth:2021bsp}. 
Over the past two decades, significant progress has been made in understanding the properties of nuclear matter around saturation and at sub-saturation densities.
However, large uncertainties still remain regarding the properties of nuclear matter at higher densities, particularly the stiffness of the EoS and the density dependence of the symmetry energy
\cite{Sorensen:2023zkk}. 
The heavy-ion collisions (HICs) experiment based on the heavy-ion accelerator is the only way to produce high-density nuclear matter in the current terrestrial experiments, and it is also an effective way to study the EoS of high-density nuclear matter \cite{Aichelin:1991xy,Hartnack:1997ez}. Due to the HICs occurring in a transient, high-temperature, and high-density environment, it is currently impossible to directly probe the microscopic evolution during the collision process in HICs experiments \cite{Bleicher:2022kcu}. Therefore, simulating the dynamical evolution of HICs within microscopic transport models, and comparing the final state observables with experimental data, has become an effective way to extract information on the nuclear EoS \cite{Bleicher:2022kcu,Xu:2019hqg,TMEP:2022xjg,TMEP:2023ifw, Cozma:2024cwc}. In transport models, the initialization of colliding nuclei is a crucial prerequisite for accurately describing the reaction dynamics, as the construction of nuclear density distributions has a significant impact on subsequent evolution and observables \cite{TMEP:2016tup,Yang:2021gwa, Sinha:2023jas, Xu:2026vnl}.
Among the widely used transport models, quantum molecular dynamics (QMD)-type models have achieved remarkable success in describing HICs from the Fermi energy region up to RHIC energies \cite{Aichelin:1991xy,TMEP:2022xjg}.

In recent years, exotic nuclear structure features, such as nuclear deformation \cite{Jia:2021tzt,Giacalone:2021udy,Jia:2022qgl, Ryssens:2023fkv,Ye:2024slx}, neutron skin thickness \cite{Xu:2021vpn,Liu:2022kvz, Giacalone:2023cet,Liu:2023qeq, Ding:2024xxu,Li:2025mox}, and $\alpha$-clustering \cite{Broniowski:2013dia, He:2014iqa, Zhang:2020hww,Wang:2021ghq,Menon:2025apq}, have emerged as a hot research topic in RHIC \cite{Li:2019kkh,Zhao:2024lpc,STAR:2024wgy,Kincses:2025iaf}. In simulations of RHIC, these structural features can be implemented by generating the coordinates of nucleons in the projectile or target nucleus according to prescribed nuclear density profiles \cite{Wang:1991hta, Gyulassy:1994ew}. However, in transport models at intermediate and low energies, such as the QMD-type, each nucleon is represented by a finite-width Gaussian wave packet, and the nuclear density distribution arises from the superposition of all wave packets \cite{Bass:1998ca,Bleicher:1999xi,TMEP:2017mex,Steinheimer:2024eha}. As a result, the challenge of reproducing non-standard or exotic radial density profiles during the initialization phase remains long-standing in QMD-type models \cite{TMEP:2016tup}. To address this issue, we propose a general and portable initialization scheme based on Fourier series expansion (FSE) filtering. This method is integrated into the UrQMD model to simulate $^{96}$Ru+$^{96}$Ru collisions at $E_{\mathrm{lab}} = 1500~\mathrm{MeV/nucleon}$. The results are then compared with those from the standard Woods–Saxon (WS) initialization to validate the effectiveness of this approach and to assess its impact on physical observables.

The structure of this paper is organized as follows. 
In Sec. \ref{sec2}, we briefly introduce the initialization of QMD-type transport models, the UrQMD model, and the newly developed filtering method. 
In Sec. \ref{sec3}, the effectiveness of the proposed method is demonstrated, and the simulated results based on the UrQMD model are presented. 
Finally, conclusions and outlook are given in Sec. \ref{sec4}.

\section{Methodology}\label{sec2}
\subsection{Initialization of QMD-type transport models}

In QMD-type transport models, nucleons are usually represented by the Gaussian wave packet with a finite width \cite{Aichelin:1991xy}:
\begin{equation}\label{GSE}
\psi_{i}(\vec{r})=\frac{1}{(2 \pi \sigma_r^{2} )^{3 / 4}} \exp \left [ -\frac{\left(\vec{r}-\vec{r}_{i}\right)^{2}}{4 \sigma_r^{2}}+\frac{i \vec{p}_{i} \cdot \vec{r}}{\hbar} \right ],
\end{equation}
where $\sigma_r^2$ is the width of the Gaussian wave packet, and $\vec{r}_{i}$ and $\vec{p}_{i}$ denote the coordinate and momentum of the center of the $i$-th nucleon wave packet, respectively.
By performing a Wigner transformation of Eq.~\ref{GSE}, the phase-space distribution function of the $i$-th nucleon can be obtained as
\begin{equation}
f_{i}(\vec{r}, \vec{p})=\frac{1}{(\pi \hbar)^{3}} \exp \left[\frac{-\left(\vec{r}-\vec{r}_{i}\right)^{2}}{2 \sigma_{r}^{2}}\right] \exp \left[\frac{-\left(\vec{p}-\vec{p}_{i}\right)^{2}}{2 \sigma_{p}^{2}}\right].
\end{equation}
The widths in coordinate and momentum space satisfy the Heisenberg uncertainty principle, i.e., $\sigma_{r}\cdot \sigma_{p}\ge  \hbar/2$. 
Next, by integrating the phase space distribution over the entire space, the density distribution of the nuclear system can be obtained as
\begin{eqnarray}\label{rho}
\rho(\vec{r}) &=& \frac{1}{(2\pi\sigma_r^2)^{3/2}} \sum_i \exp{\left[ -\frac{(\vec{r} - \vec{r}_i)^2}{2\sigma_r^2} \right]}.
\end{eqnarray}

For the initialization of QMD-type models, the Monte Carlo method is first used to sample nucleon spatial coordinates within a spherical volume of radius $R$. 
Then, the density distribution is determined by Eq.~\ref{rho}, and the WS distribution is further used to filter the sampled nucleons. 
The WS distribution is a classical model describing the nuclear density profile, expressed as
\begin{equation}\label{wsd}
\rho^\mathrm{WS}(r) = \frac{\rho_{0}}{1+\exp\left(\frac{r-C}{a}\right)},
\end{equation}
where $\rho_0=0.16~\mathrm{fm}^{-3}$ is the saturation density of nuclear matter, $C$ is the half-density radius, and $a$ is the surface diffuseness parameter \cite{Myers:1983seb,Gupta:2009zzb, Sun:2009wf}.
For illustrative purposes, when a one-dimensional case is discussed in the following text, the saturation density is set as $\rho_0 = 0.16~\mathrm{fm^{-1}}$, corresponding to a linear density.
After the coordinates of all nucleons are determined, nucleon momenta are randomly sampled within the range $[0,p_{F}]$, where the local Fermi momentum is defined as $p_{F}=(3\pi^{2}\rho(\vec{r}))^{1/3}$. 

The next step is to calculate the binding energy of the initialized nucleus. If the calculated binding energy per nucleon is in the range of 6.5-8.5 MeV, the initialized nucleus will be used. Otherwise, the momenta will be resampled. It is known that the binding energy per nucleon of $^{96}$Ru is about 8.6 MeV \cite{Wang:2012eof}.
To examine the effect of initial binding energy on observables, we have further changed the binding energy cut to 8.5-8.7 MeV and compared the corresponding results. It is found that the directed flow and elliptic flow of free protons under different binding energy cuts are almost identical. This is understandable because, compared with the beam energy of 1500 MeV/nucleon, the change in the binding energy constraint in the initialization is negligible.
In standard QMD-type models, the many-body wave function is approximated as a direct product of single-particle wave packets without explicit antisymmetrization. As a result, fermionic correlations are not fully incorporated, and the variational energy minimum may not represent a physically realistic ground state, as discussed in Ref. \cite{Maruyama:2012bi}. 
It should be noted, however, that extended approaches, such as constrained molecular dynamics (CoMD) \cite{Papa:2000ef}, introduce phase-space occupation constraints to partially restore fermionic features. Therefore, constructing a fully self-consistent nuclear ground state remains a nontrivial task in most QMD-based transport models. \cite{Aichelin:1991xy, Hartnack:1997ez,TMEP:2016tup}.
In contrast, models such as antisymmetrized molecular dynamics (AMD), fermionic molecular dynamics (FMD), and extended quantum molecular dynamics (EQMD) usually employ methods such as “frictional cooling” to construct initial nuclei that are closer to the true ground state, which is more suitable for studies of low energy heavy-ion reactions  \cite{Feldmeier:1989st,He:2014iqa, Myo:2023alz,Shi:2024scb}.
In the present work, the beam energy is 1500 MeV/nucleon, which is much higher than the nuclear binding energy. It is reasonable not to further consider a strict lowest energy state treatment in the present study.


\subsection{UrQMD model}
In this work, the UrQMD model will be used as the benchmark model for testing. 
And it is a typical transport model used for microscopic many-body non-equilibrium dynamics \cite{Sombun:2018yqh,Li:2021sdc,Neidig:2021bal,Li:2022wvu}. 
The coordinates and momentum of each nucleon are propagated using Hamilton’s equations of motion \cite{Bleicher:1999xi, Li:2011zzp}:
\begin{equation}
\dot{\vec{p} }_i = -\frac{\partial \left \langle H \right \rangle }{\partial \vec{r_i} }, \quad \dot{\vec{r} }_i = \frac{\partial \left \langle H \right \rangle }{\partial \vec{p_i} },
\end{equation}
where $\langle H \rangle$ denotes the total Hamiltonian of the system. It consists of the kinetic energy $T$ and the effective interaction potential $U$, including the Skyrme interaction $U_\mathrm{\rho}$, the Coulomb interaction $U_\mathrm{Coul}$, and the momentum-dependent interaction $U_\mathrm{md}$,
\begin{equation}
\langle H\rangle = T + U_\mathrm{\rho} + U_\mathrm{Coul} + U_\mathrm{md}.\
\end{equation}
To study the HICs at intermediate energies, the Skyrme energy density functional was introduced \cite{Wang:2013wca,Wang:2016yti,Wang:2020vwb,Xiao:2023pqs,Li:2018bus}. The local and momentum-dependent potential energies can be written as $U_\mathrm{\rho, md} = \int u_\mathrm{\rho, md} \mathrm{d}\vec{r}$, 
\begin{equation*}
\begin{aligned}
u_{\rho} &= \frac{\alpha}{2}\frac{\rho^{2}}{\rho_{0}}
+ \frac{\beta}{\gamma+1}\frac{\rho^{\gamma+1}}{\rho_{0}^{\gamma}} \\
&\quad + \frac{g_{\mathrm{sur}}}{2\rho_{0}}(\nabla\rho)^{2}
+ \frac{g_{\mathrm{sur,iso}}}{2\rho_{0}}
    \left[\nabla\left(\rho_{\mathrm{n}}-\rho_{\mathrm{p}}\right)\right]^{2} \\
&\quad + \left[
A_{\mathrm{sym}}\left(\frac{\rho}{\rho_0}\right)
+ B_{\mathrm{sym}}\left(\frac{\rho}{\rho_0}\right)^{\eta}
+ C_{\mathrm{sym}}\left(\frac{\rho}{\rho_0}\right)^{5/3}
\right]\delta^2\rho,
\end{aligned}
\end{equation*}
where, $\delta = (\rho_\mathrm{n} - \rho_\mathrm{p}) / (\rho_\mathrm{n} + \rho_\mathrm{p})$ denotes the isospin asymmetry, while $\rho_\mathrm{p}$ and $\rho_\mathrm{n}$ represent the proton and neutron densities, respectively. 
And the momentum-dependent part is written as
\begin{equation}
u_{\mathrm{md}} = t_{\mathrm{md}} \ln^2 \left[ 1 + a_{\mathrm{md}} (\vec{p}_i - \vec{p}_j)^2 \right] \frac{\rho}{\rho_0}.
\end{equation}

In this work, to further verify the effectiveness and robustness of the initialization methods under different EoS, the soft momentum-dependent (SM, $K_{0} = 200~\mathrm{MeV}$) and medium-hard momentum-dependent (mHM, $K_{0} = 280~\mathrm{MeV}$) EoS are employed.
The corresponding parameter sets are listed in Tab. \ref{tab11}, while all other parameters remain consistent \cite{Huth:2021bsp,LeFevre:2015paj, Li:2021ikk, Xu:2021aij, Tsang:2023vhh}, 
$g_{\mathrm{sur}} = 19.5~\mathrm{MeV \cdot fm}^2$,
$g_{\mathrm{sur,iso}} = -11.3~\mathrm{MeV \cdot fm}^2$,
$A_{\mathrm{sym}} = 20.4~\mathrm{MeV}$,
$B_{\mathrm{sym}} = 10.8~\mathrm{MeV}$,
$C_{\mathrm{sym}} = -9.3~\mathrm{MeV}$,
$\eta = 1.3$,
$t_{\mathrm{md}} = 1.57~\mathrm{MeV}$,
and $a_{\mathrm{md}} = 500~(\mathrm{GeV}/c)^{-2}$.

\begin{table}[htbp!]
\centering
\renewcommand{\arraystretch}{2}
\begin{tabular}{c @{\hskip 0.3cm} c @{\hskip 0.3cm} c @{\hskip 0.3cm} c @{\hskip 0.3cm} c  @{\hskip 0.3cm} c}
\hline\hline
EoS & $K_{0}$~(MeV) & $L$~(MeV) & $\alpha$~(MeV) & $\beta$~(MeV) & $\gamma$ \\
\hline
SM & 200 & 80.95 & -390 & 320 & 1.14\\
mHM & 280 & 80.95 & -162 &  97 & 1.57\\
\hline\hline
\end{tabular}
\caption{Parameter sets of the nuclear EoS used in this work.}
\label{tab11}
\end{table}

\subsection{Filtering method}

To ensure that exotic nuclear density distributions can be accurately obtained through random sampling of nucleon coordinates $\vec{r_i}$ during the initialization process of QMD-type transport models, we propose a new nucleon coordinate filtering method, the basic principle of which is described below.

For clarity, we first illustrate the procedure in the one-dimensional case along the $z$ direction as an example. The target distribution is assumed to be as follows:
\begin{equation}\label{MBF}
f(z) = a + a_1 \cdot \cos(\frac{\pi}{l} z),
\end{equation}
where $a$ and $a_{1}$ are coefficients, and $l$ represents the length parameter. To achieve this, one needs to determine the distribution of the wave-packet centroids, denoted as $g(z_i)$, which satisfies
\begin{equation}\label{MBF2}
f(z) = \frac{1}{(2\pi\sigma_{r}^2)^{1/2}} \sum_i \int_{-\infty}^{+\infty} g(z_i) \exp\left[-\frac{(z-z_i)^2}{2\sigma_{r}^2}\right] dz_i,
\end{equation}
where $z_i$ represents the center of the Gaussian wave packet. The corresponding filtering function can be obtained as
\begin{equation}\label{MBF4}
g(z_i) = a + a_1 \cdot \exp\left(\frac{\pi^{2}\sigma_{r}^{2}}{2l^2}\right) \cos\left(\frac{\pi z_i}{l}\right),
\end{equation}
if the centroids are uniformly distributed, $g(z_i)$ is constant (e.g., $g(z_i)=1$), and the resulting $f(z)$ is constant, failing to reproduce the target distribution. In contrast, using the filtering function in Eq.~\ref{MBF4} correctly reproduces the desired $f(z)$.

For the one-dimensional WS distribution, the following formula can be expressed by Fourier series expansion, and with the increase of cosine terms, the expansion gradually converges to the original function, 
\begin{equation}\label{MBF3}
\rho^\mathrm{WS}(z) = a_0 + \sum_{j=1}^{n} a_j \cos(c_j z),
\end{equation}
where $a_{0}$ and $a_{j}$ denote the zeroth- and $j$th-order expansion coefficients, respectively,
\begin{align}
a_0 &= \frac{1}{l}\int_{0}^{l}\rho^\mathrm{WS}(z) \, dz, \\
a_j &= \frac{2}{l}\int_{0}^{l}\rho^\mathrm{WS}(z)\cos(c_j z) \, dz, \\
c_j &= j\pi/l.
\end{align}

To obtain the WS density distribution expressed in Eq.~\ref{wsd}, the filtering function is expressed as
\begin{equation}
g(z_i) = a_{0} + \sum_{j=1}^{n} a_{j} \exp\left( \frac{c_j^{2} \sigma_{r}^{2}}{2} \right) \cos(c_j z_i),
\end{equation}
with its maximum value given by
\begin{equation}
g_{\mathrm{max}} = a_{0} + \sum_{j=1}^{n} \left | a_{j} \right |  \exp\left( \frac{c_j^{2} \sigma_{r}^{2}}{2} \right).
\end{equation}
During the random sampling procedure, only the region satisfying $0 < g(z_i) < g_{\mathrm{max}}$ is retained, such that the filtering function remains strictly positive.
Then, random coordinate points are uniformly selected along the $z$ direction, their corresponding $g(z_i)$ values are calculated, and further compared with random numbers within the interval $\left[0, g_{\mathrm{max}}\right]$, if $g(z_i)$ is greater than the random number, then the coordinate is retained, otherwise discard the coordinate. 
This process is repeated until enough nucleon coordinates are obtained. 
Finally, the density distribution of the nucleus is calculated by
\begin{equation}\label{rhoo}
\rho(z) = \frac{1}{(2\pi\sigma_{r}^2)^{1/2}} \sum_{i}^{} \exp\left[ -\frac{(z - z_{i})^2}{2\sigma_{r}^2} \right].
\end{equation}

\begin{figure}[t]
\centering
\includegraphics[width=1.0\linewidth]{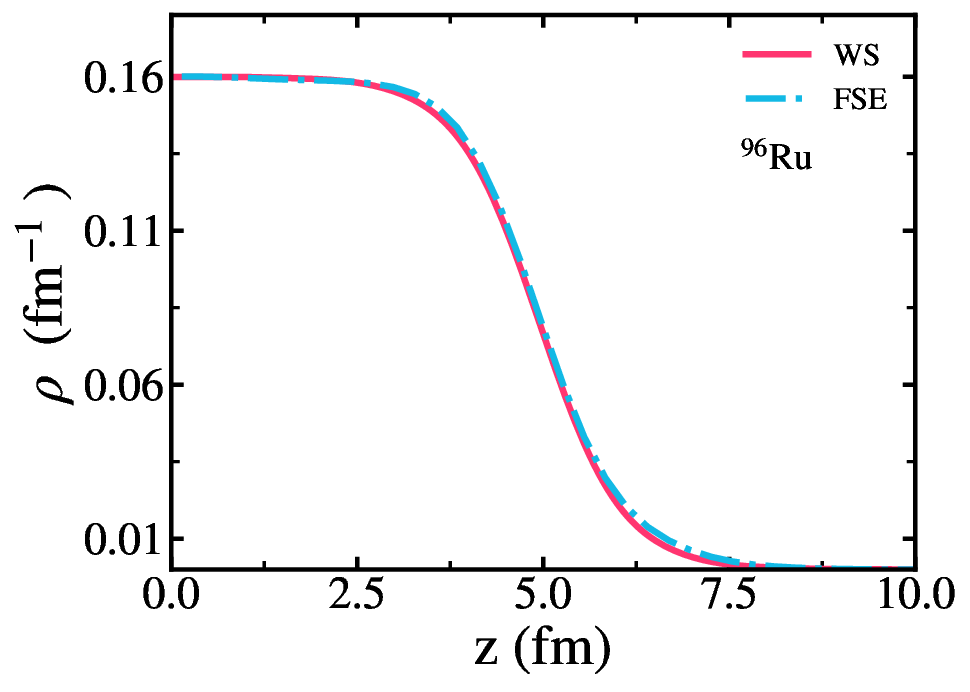}
\caption{(Color online) Comparison of the density distributions obtained from the WS function Eq.~\ref{MBF3} approximated by a Fourier series truncated at twenty terms and from the FSE filtering method in the one-dimensional ($z$) case.}
\label{fig.1}
\end{figure}

To verify the validity of the above filtering method, Fig.~\ref{fig.1} presents a comparison between the density distributions obtained from the WS function (Eq.~\ref{MBF3}) approximated by a twenty-term Fourier series with $l=16~\mathrm{fm}$ (red solid line) and those derived using the FSE filtering method with Eq.~\ref{rhoo} (blue dash-dotted line). 
It can be seen that the density distribution after FSE processing shows excellent agreement with the original WS distribution, indicating that the new filtering method can accurately reproduce the objective density distribution in the one-dimensional case.

However, in realistic nuclear physics applications, it is essential to model the full three-dimensional nuclear density distribution. Accordingly, Eq.~\ref{MBF2} is generalized to
\begin{equation}
f(\vec{r}) = \frac{1}{(2\pi\sigma_{r}^2)^{3/2}} \sum_i \iiint_{-\infty}^{+\infty} g(\vec{r_{i}}) \exp \left [ -\frac{(\vec{r} -\vec{r_{i}})^2}{2\sigma_{r}^2}\right ]  d^3\vec{r_{i}},
\end{equation}
and for even functions like Eq. \ref{MBF3}, the corresponding filtering function takes the form
\begin{equation}
g(\vec{r_{i}})=a_0+\sum_{j=1}^na_j \exp\left( \frac{c_j^2\sigma_{r}^2}{2} \right)\cos(c_j ||\vec{r_{i}}||),
\end{equation}
where $||\vec{r_{i}}||$ represents the magnitude of the position vector $\vec{r_i}$.

Furthermore, when dealing with non-analytic exotic nuclear density distributions, especially discrete data-type distributions from experiments or theoretical calculations, the traditional FSE method encounters new difficulties. 
On the one hand, it is difficult to directly construct explicit forms of the objective function and its filtering function. 
On the other hand, it is challenging to analytically solve the expansion coefficients. 
To solve these problems, while keeping the Fourier expansion framework, numerical integration is employed instead of analytical integration to calculate the expansion coefficients,
\begin{align}
a_0 &= \frac{1}{l}\sum_{k=0}^{n} \rho(r_k)\Delta r_k, \\
a_j &= \frac{2}{l}\sum_{k=0}^{n} \rho(r_k)\cos\left(\frac{j\pi r_k}{l}\right) \Delta r_k,
\end{align}
where $\rho(r_k)$ represents the initial nuclear density at radius $r_k$, and $\Delta r_k$ is the discrete sampling interval. 

\section{Results and discussion}\label{sec3}
\subsection{Reconstruction of the objective density distribution}

\begin{figure}[t]
\centering
\includegraphics[width=0.85\linewidth]{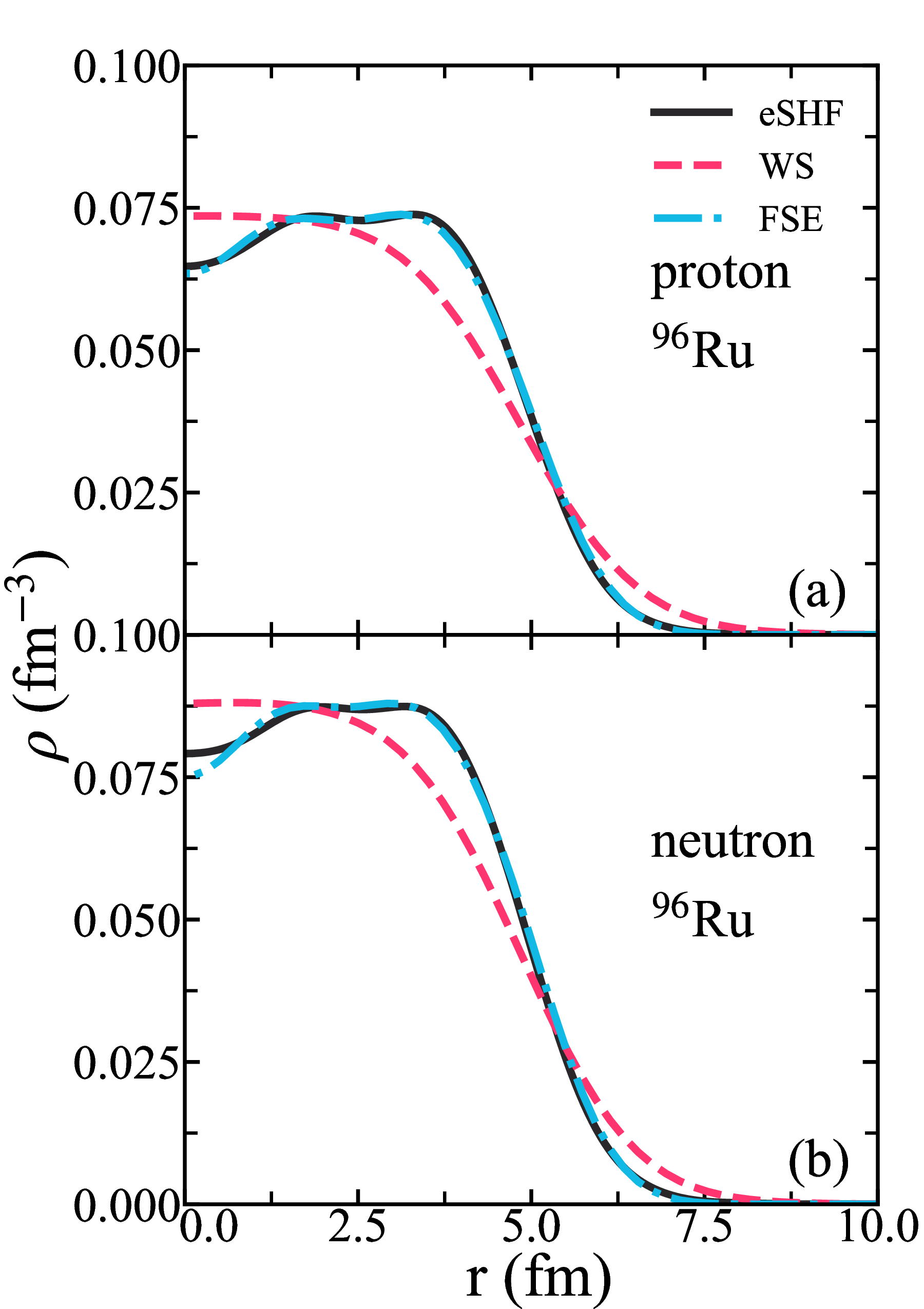}
\caption{(Color online) Proton (top) and neutron (bottom) density distributions of $^{96}$Ru. 
The eSHF denotes the objective density distributions taken from Ref. \cite{Li:2019kkh}, while the WS and the FSE represent results obtained using Eqs. \ref{wsd} and \ref{fes_3D} as filtering functions, respectively.}
\label{fig.2}
\end{figure}

Firstly, to evaluate the capability of the new method in reconstructing the target density distribution, the case of $^{96}$Ru+$^{96}$Ru is taken as an example.
The proton and neutron density distributions of $^{96}$Ru exhibiting a bubble structure, calculated using the extended Skyrme–Hartree–Fock (eSHF) method in Ref.~\cite{Li:2019kkh}, are adopted as the target densities.
These data are discretely resampled at a radial interval of $\Delta r = 0.02$ fm over the range $r \in [0, 16]$ fm, yielding $n = 800$ uniformly spaced grid points.
Here, $l = 16~\mathrm{fm}$, and the Fourier series expansion (FSE) is truncated at 20 terms to obtain the reconstructed density distribution.
Based on this, the corresponding filtering function can be expressed as
\begin{equation}\label{fes_3D}
g(\vec{r_{i}}) = a_0 + \sum_{j=1}^{20} a_j \cdot \exp \left( \frac{j^2 \pi ^2\sigma_{r}^2}{512} \right) \cos\left( \frac{j\pi}{16} ||\vec{r_{i}}|| \right).
\end{equation}
It should be emphasized that this new method requires about 1.5 times longer initialization than the traditional approach, but this increase can be negligible compared to the total dynamical evolution time. 

Then, the proton and neutron density distributions of $^{96}$Ru are obtained by performing the initialization sampling using different nucleon coordinate filtering methods, the results are shown in Fig.~\ref{fig.2}. 
The black solid line represents the objective density distributions from Ref. \cite{Li:2019kkh}, while the red dashed line and the blue dash-dotted line correspond to the results of the WS filtering method and the FSE filtering method, which are calculated by using Eqs. \ref{wsd} and \ref{fes_3D}, respectively. 
It can be seen that the results obtained with the WS method deviate significantly from the target density distribution, whereas the FSE filtering method accurately reproduces the desired proton and neutron density profiles and successfully captures the exotic structural features.

\begin{figure}[!htb]
\centering
\includegraphics[width=\linewidth]{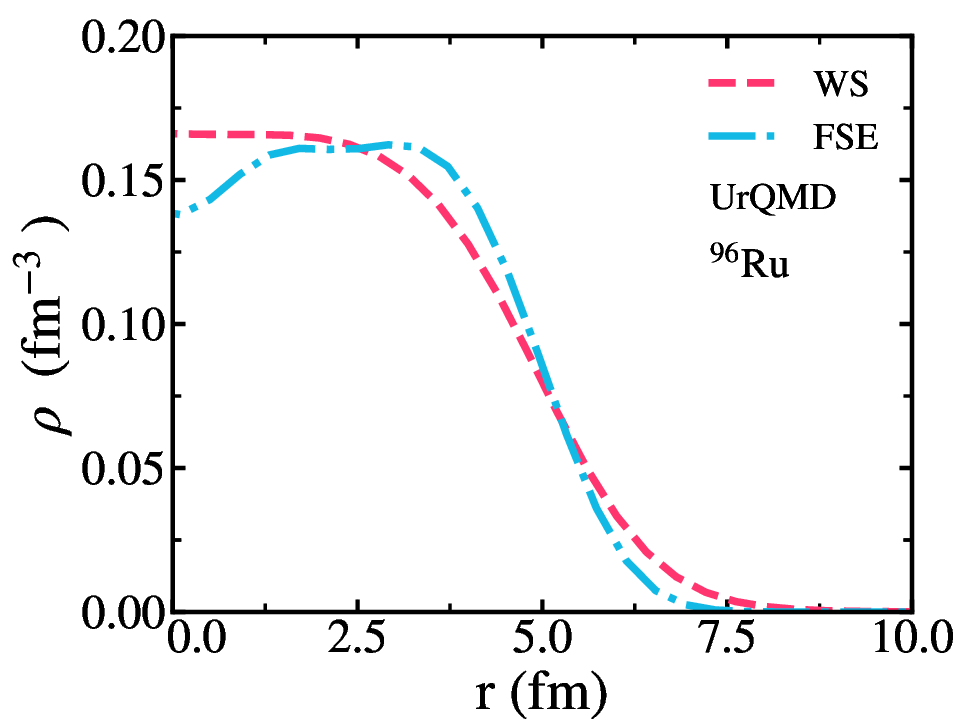}
\caption{(Color online) Initial density distribution of $^{96}$Ru obtained from the UrQMD model with WS and FSE filtering method.}
\label{fig.3}
\end{figure}

\subsection{Applying the FSE method to the UrQMD model}

\begin{figure*}[!htb]
\centering
\includegraphics[width=\linewidth]{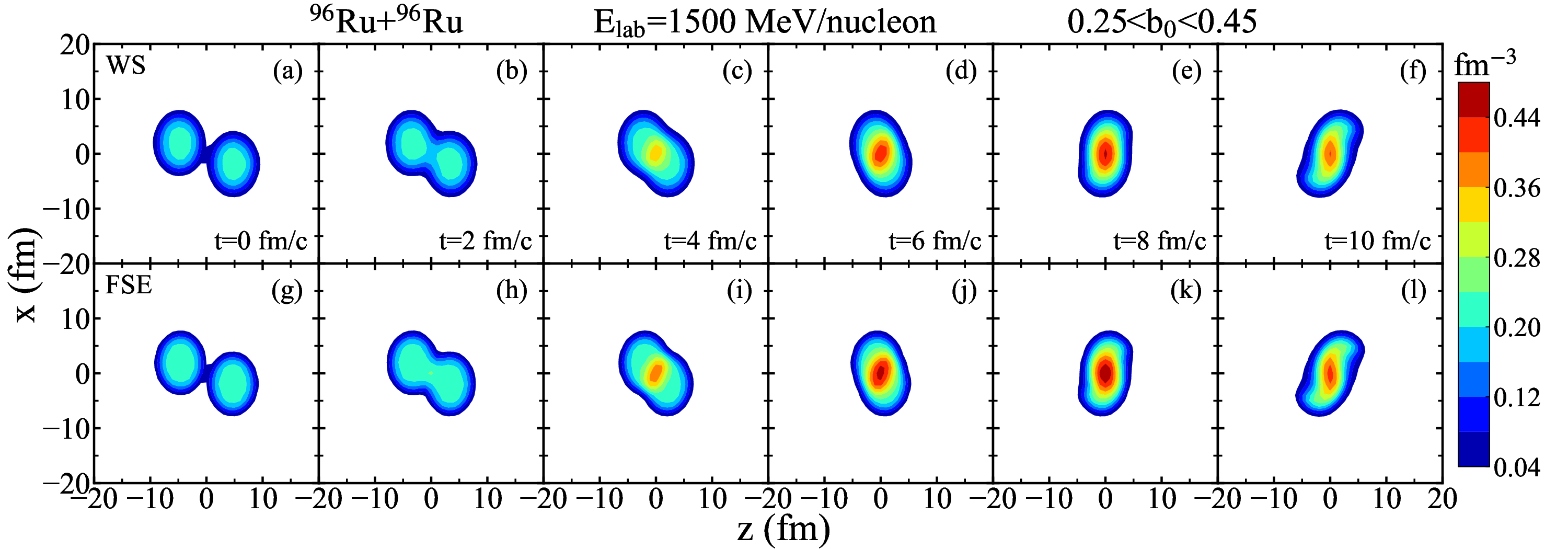}
\caption{(Color online) Time evolution of the density distributions in semi-central $^{96}$Ru+$^{96}$Ru collisions ($0.25 < b_0 < 0.45$) at a beam energy of $E_{\mathrm{lab}}=1500~\mathrm{MeV/nucleon}$, obtained within the UrQMD model using the WS and FSE filtering methods.}
\label{fig.4}
\end{figure*}

Next, taking the UrQMD model as an example, the FSE filtering method is further incorporated into the transport model.  
Figure~\ref{fig.3} presents the initial density distributions of $^{96}$Ru, which are output from the UrQMD model with WS (red solid
line) and FSE (blue dash-dotted line) filtering methods. 
The desired bubble-like structure of $^{96}$Ru can be successfully reproduced by adopting the new FSE filtering method. 

In studies which combine the transport model simulations with HICs experiments to constrain nuclear structure, generating projectile and target nuclei with exotic nucleon density distributions or structural effects during initialization is only the first step.
The crucial prerequisite for reliable results is that the sampled nuclei maintain their internal structures and shapes stably. 
Thus, the time evolution of the density distributions and the root-mean-square (RMS) radius of $^{96}$Ru are further checked. 
It is found that the nuclei obtained by the FSE method can better preserve the exotic structures, i.e., the desired objective density distributions with a bubble-like structure in the central region, until the geometrical overlap time ($\sim6.2$ fm/c) of $^{96}$Ru+$^{96}$Ru collisions at $E_{\mathrm{lab}} = 1500~\mathrm{MeV/nucleon}$ is reached. 
The complete geometrical overlap time of the colliding nuclei is estimated with the same formulism as Ref. \cite{Reichert:2024ayg}.
And at 10 fm/c, the RMS radius of the FSE method increases about 1.8\% while that of the WS method increases by 1.1\%, due to the inherent fluctuations of the QMD-like model \cite{TMEP:2021ljz}, these variations are within acceptable limits, and do not threaten the following conclusions. 
These results imply that in the UrQMD model, nuclei initialized using the FSE filtering method not only capture the desired structural effects but also remain stable until the system reaches maximum geometric overlap.

\subsection{Analysis of model simulation results}

Last, the effects of the initial nuclear structure on the non-equilibrium dynamical process and the observables are explored. Figure~\ref{fig.4} shows the time evolution of the density distributions of the system obtained with the two filtering methods.
It is evident that the different initial density profiles of the colliding nuclei lead to pronounced differences in the system’s density distributions, particularly in the central region, where the highest densities are achieved during the collision.
Compared with collisions initialized using the WS method, those initialized with the FSE method, i.e., featuring bubble-like nuclear structures, more readily attain a higher maximum central density. 
Combined with Fig.~\ref{fig.3}, one can see that nuclei generated with the FSE method exhibit broader density profiles at the initial stage, which gradually narrow with time, leading to an increase in the central density. 
This results in a higher central density in the collision system compared to that obtained with the WS initialization.

\begin{figure}[b]
\centering
\includegraphics[width=1.0\linewidth]{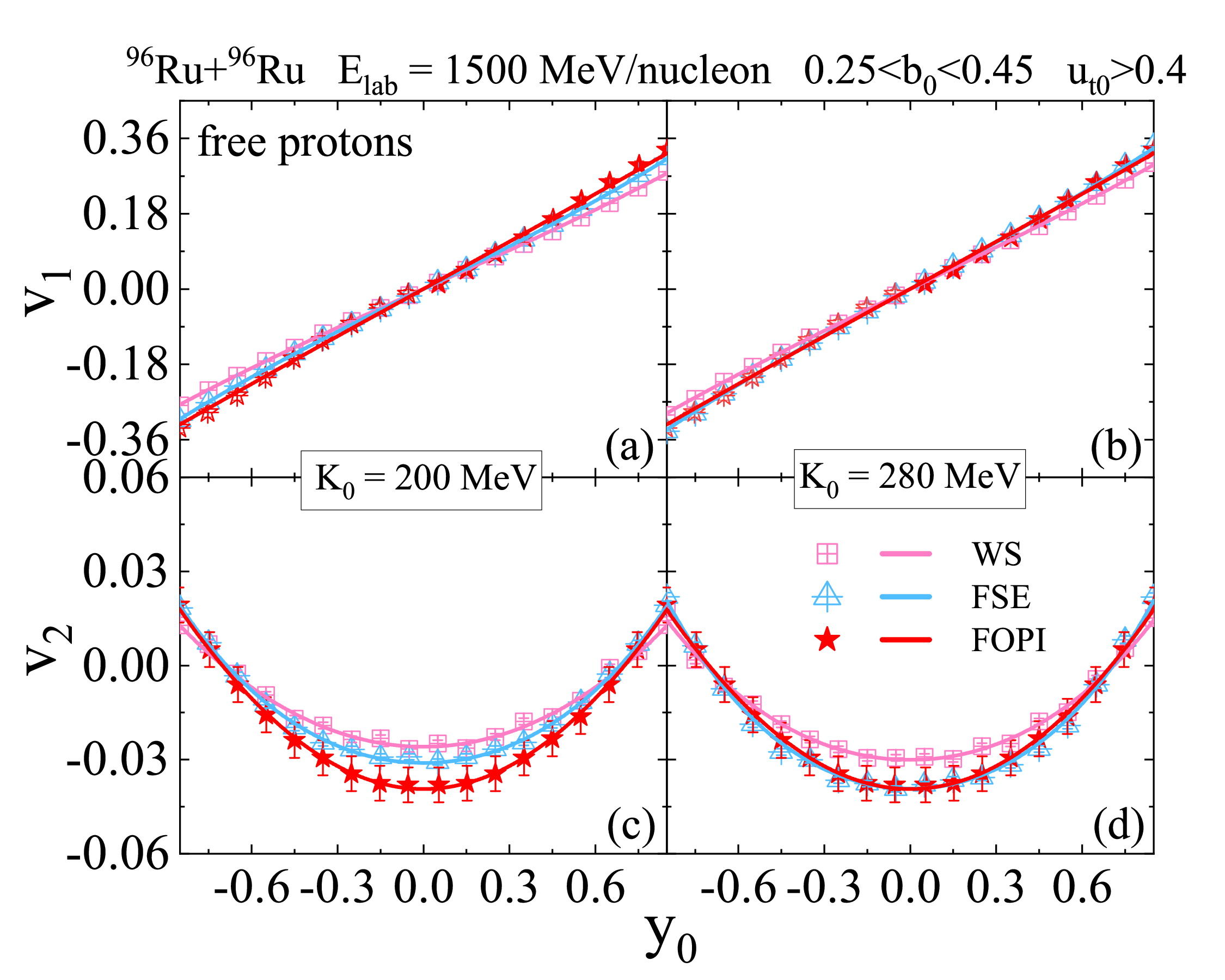}
\caption{(Color online) Reduced rapidity dependence of the directed flow $v_{1}$ (top) and elliptic flow $v_{2}$ (bottom) of free protons in semi-central $^{96}$Ru+$^{96}$Ru collisions ($0.25 < b_{0} < 0.45$) at $E_{\mathrm{lab}} = 1500~\mathrm{MeV/nucleon}$ with transverse momentum $u_{t0} > 0.4$. Left panels correspond to calculations with the incompressibility coefficient $K_{0} = 200~\mathrm{MeV}$, while right panels show the results calculated with $K_{0} = 280~\mathrm{MeV}$. The experimental data are taken from FOPI Collaboration~\cite{FOPI:2011aa}. The lines are fits to the calculated results and data (see text).}
\label{fig.5}
\end{figure}

A large body of research has demonstrated that collective flow is a sensitive observable for constraining the nuclear EoS.
In particular, the directed flow ($v_{1}=\langle p_x/\sqrt{p_x^2+p_y^2}\rangle$) and elliptic flow ($v_{2}=\langle(p_x^2-p_y^2)/(p_x^2+p_y^2)\rangle$) at mid-rapidity are widely used to probe the stiffness of the high-density nuclear EoS \cite{Danielewicz:2002pu,FOPI:2011aa,Wang:2018hsw}. 
Figure.~\ref{fig.5} shows the reduced-rapidity ($y_{0}$) dependence of the $v_{1}$ and $v_{2}$ of free protons in semi-central $^{96}$Ru+$^{96}$Ru collisions ($0.25 < b_0 < 0.45$) at $E_{\mathrm{lab}}=1500~\mathrm{MeV/nucleon}$. 
The definition of the reduced-rapidity and the transverse momentum cut are the same as that of the FOPI experiment \cite{FOPI:2011aa}. 
The left and right panels correspond to calculations with incompressibility coefficients $K_{0}$=200 and 280 MeV, respectively. 
The red stars denote the FOPI experimental data (open stars indicate anti-symmetrized data) \cite{FOPI:2011aa}. 
And the results obtained using the WS and FSE initialization methods are represented by open pink squares and open blue triangles, respectively, while the solid lines correspond to polynomial fits of the form $v_{1}(y_{0}) = v_{11}y_{0} + v_{13}y_{0}^{3} + c$ and $v_{2}(y_{0}) = v_{20} + v_{22}y_{0}^{2} + v_{24}y_{0}^{4}$ within $|y_{0}| < 0.85$.

\begin{figure}[t]
\centering
\includegraphics[width=1.0\linewidth]{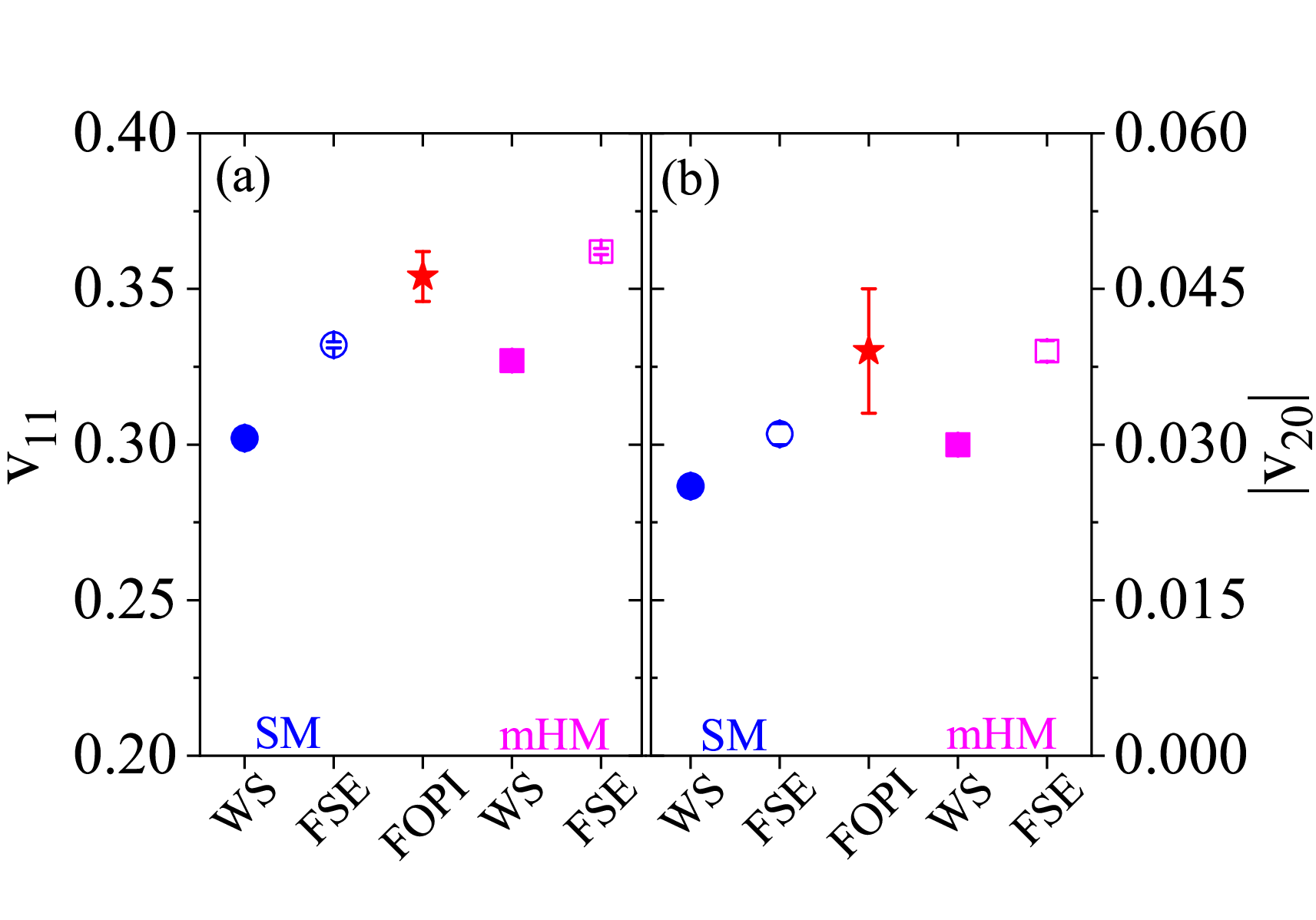}
\caption{(Color online) The $v_{1}$ slope and $v_{2}$ at mid-rapidity for free protons obtained from fitting to the calculated results and data shown in Fig. \ref{fig.5}.}
\label{fig.6}
\end{figure}

As shown in Fig.~\ref{fig.5}, the calculated $v_{1}$ exhibits good agreement with the FOPI data in the whole inspected rapidity range for both EoS parameterizations and initialization methods.
As for $v_{2}$, due to the EoS with $K_{0}=280$ MeV generating a stronger repulsive interaction than the softer $K_{0}=200$ MeV case, and the FSE initialization method yields a higher system density than the WS scheme, the $|v_{2}|$ values obtained with the stiffer EoS or FSE initialization are larger and closer to the experimental data. 
This effect is most evident at mid-rapidity, where the sensitivity to the high-density nuclear EoS is strongest. 

To better quantify the impact of the initialization method on collective flow, Fig.~\ref{fig.6} shows the $v_{11}$ and $|v_{20}|$ at mid-rapidity extracted from fitting to the calculated results and experimental data shown in Fig.~\ref{fig.5} are summarized in Fig.~\ref{fig.6}. 
It can be seen that with the WS initialization method, both the values of $v_{11}$ and $|v_{20}|$ calculated using mHM EoS (solid squares), are larger than those calculated using SM EoS (solid circles) and closer to the data (stars),but they still fail to fully reproduce the data. 
For the new FSE initialization method, the data of $v_{11}$ falls between the results calculated using SM EoS (open circles) and mHM EoS (open squares), while the data of $|v_{20}|$ is larger than the result calculated using the SM EoS (open circles) and approximately equal to the result calculated using the mHM EoS (open squares). 
Recently, several studies indicated that a gradual hardening of the EoS of hot and dense nuclear matter occurs as its density increases in the reactions with higher beam energies, and a consecutive softening with the QCD phase transition \cite{Li:2022iil,OmanaKuttan:2022aml,Li:2025iqq, Chen:2025ijh}.
Our results indicate that at the investigated energy, the FOPI collective flow data of $^{96}$Ru+$^{96}$Ru collisions favor a relatively stiff EoS ($K_{0}>$280 MeV) when using the WS initialization method, and an EoS of $K_{0}=200-280~\mathrm{MeV}$ when using the FSE initialization method.

\section{SUMMARY and Outlook}\label{sec4}

This work focuses on the initialization of projectile and target nuclei in QMD–type transport models, specifically the sampling of nucleon coordinates and the corresponding density distributions.
A filtering method based on FSE is proposed, in which the filtering function for nucleon coordinates is obtained by performing FSE of the desired nuclear density distribution. 
Then, within the UrQMD framework, the $^{96}$Ru nucleus with a bubble-like structure is initialized using the proposed FSE method, and the resulting density distributions are shown to reproduce the desired profiles more accurately than those obtained with the conventional WS initialization.

Furthermore, to validate the effectiveness of this new method and evaluate the impact of the initial density distribution on the observables, the $^{96}$Ru+$^{96}$Ru collisions at $E_{\mathrm{lab}}=1500~\mathrm{MeV/nucleon}$ are performed with the UrQMD model under the WS and FSE initialization methods.   
It is found that the maximum density of the system is increased and the collective flow effect is enhanced by bubble-like density distribution. As a consequence, the extraction of EoS information exhibits dependence on the assumed initial density distribution, with a relatively stiff EoS with $K_{0}>280~\mathrm{MeV}$ is favored when adopting the WS-type density profile, whereas a softer EoS with $K_{0}=200-280~\mathrm{MeV}$ is supported when a bubble-like density distribution is employed. 
These findings indicate that, within the microscopic transport models, the FSE filtering method can be used to sample nuclei with exotic density profiles, thereby enabling improved investigations of nuclear structure and high-density nuclear EoS issues through HICs.

In this study, it is shown that the initialization density distribution of colliding nuclei directly influences the subsequent dynamical evolution of the density, thereby affecting flow observables. 
It is interesting to explore the collisions of nuclei with higher isospin asymmetry, which exhibit richer structural and density distribution characteristics. The related investigations will be helpful for constraining the EoS of isospin asymmetry nuclear matter, and for understanding of the formation and evolution of compact stars. 
Furthermore, the effects of exotic nuclear density distributions under different beam energies and reaction systems are also worth investigating.
More systematic studies that incorporate these effects are necessary, and related work is ongoing.

\section*{acknowledgments}
The study is supported in part by the National Natural Science Foundation of China (Nos. 12335008 and 12505143), the National Key Research and Development Program of China under (No. 2023YFA1606402), the National Natural Science Foundation of Zhejiang Province (No. LQN25A050003), the Huzhou Natural Science Foundation (2024YZ28), the Scientific Research Fund of the Zhejiang Provincial Education Department (No. Y202353782), the Foundation of National Key Laboratory of Plasma Physics (Grant No. 6142A04230203). The authors are grateful to the C3S2 computing center in Huzhou University for calculation support.

\bibliography{ref.bib}
	
\end{document}